\documentclass[aps,pra,preprint,superscriptaddress,showpacs]{revtex4}
\usepackage{graphics}
\usepackage{bm}
\begin{document}

\title{
Linear response of heat conductivity of normal-superfluid interface of a polarized Fermi gas to orbital magnetic field}

\author{N. Ebrahimian}
\email{n.ebrahimian@aut.ac.ir}
\affiliation{Physics Department, Amirkabir University of Technology, Tehran 15914, Iran}
\author{M. Mehrafarin}
\email{mehrafar@aut.ac.ir} 
\affiliation{Physics Department, Amirkabir University of Technology, Tehran 15914, Iran}
\author{R. Afzali}
\email{afzali@kntu.ac.ir} 
\affiliation{Physics Department, K.N. Toosi University of Technology, Tehran 15418, Iran}

\begin{abstract}
Using perturbed Bogoliubov equations, we study the linear response to a weak orbital magnetic field of the heat conductivity of the normal-superfluid interface of a polarized Fermi gas at sufficiently low temperature. We consider the various scattering regions of the BCS regime and analytically obtain the transmission coefficients and the heat conductivity across the interface in an arbitrary weak orbital field. For a definite choice of the field, we consider various values of the scattering length in the BCS range and numerically obtain the allowed values of the average and species-imbalance chemical potentials. Thus, taking Andreev reflection into account, we describe how the heat conductivity is affected by the field and the species imbalance. In particular, we show that the additional heat conductivity due to the orbital field increases with the species imbalance, which is more noticeable at higher temperatures. Our results indicate how the heat conductivity may be controlled, which is relevant to sensitive magnetic field sensors/regulators at the interface.
\end{abstract}

\maketitle

\section{Introduction}

Recent experiments on ultracold Fermi gases \cite {Luo,Partridge,Bourdel}, such as $^{40}\text{K} $ and
$^{6}\text{Li}$ atoms, have shifted our perspective on superfluidity and confirmed earlier ideas about fermionic paring \cite{Leggett,Nozieres,Melo}.
The extent of possible experimental control in these gases has led to a large body of experimental work on fermionic pairing phenomena. Most notable
is the ability to apply a controlled magnetic field to tune the strength of the attractions \cite{Partridge,Bourdel}. In balanced mixtures of fermions, this tunability is exploited to study the crossover from a Bose-Einstein condensate (BEC) of molecules to a Bardeen-Cooper-Schrieffer (BCS) superfluid \cite{Partridge1}. The BCS regime involves a small negative scattering length wherein fermions form weakly bound Cooper pairs, while in the BEC regime the scattering length is positive and particles form tightly bound molecular pairs. However, in the case of imbalanced mixtures of spin-polarized Fermi gases \cite{Giorgini},
the study of the phase diagram is more interesting \cite{Sheehy,Parish}. In this case, many experimental \cite{Partridge1,Zwierlein,Shin,Partridge2,Ketterle} and theoretical \cite{Liu1,Bedaque,Caldas,Carlson,Mizushima,Chevy,Haque,Silva,Son,Sheehy1,Sheehy2,Radzihovsky} studies have been reported. In the experiments, a controllable number of atoms prepared in one spin state can transfer to the opposite spin state by applying a radio-frequency pulse. In these polarized systems, all atoms in the less populated spin state are found to pair up with atoms in the opposite spin state, constituting a molecular condensate that coexists with the remaining Fermi sea of the unpaired atoms. As the repulsive interaction between these atoms and the molecules increases, a phase separation occurs between the superfluid (SF) and the normal (N) phase \cite{Partridge1}. Such a phase-separation scenario had been proposed by Clogston \cite{Clogston} and Chandrasekhar \cite{Chandraskhar} long ago, who predicted the occurrence of a first-order transition from the N to the SF state. Since the existence of a N-SF interface may settle some important questions regarding polarized Fermi mixtures, serious research has been focused on this subject \cite{Silva1,Lazarides1,Lazarides2,Baur}. In particular, it has been established \cite{Lazarides1,Lazarides2} that a temperature difference between the two phases can appear as a consequence of the blockage of energy transfer across the N-SF interface. This blockage is due to a SF gap, which causes low-energy normal particles to be reflected from the interface. By studying particle scattering off the interface, the heat conductivity has been also calculated. We may mention in passing that the heat conductivity has been calculated in other polarized Fermi systems, such as the spin-polarized superfluid Fermi liquid \cite{Afzali}, where a phase separation does not occur.

In this article we examine the N-SF interface of a polarized Fermi gas in the presence of a weak orbital magnetic field, which is produced fictitiously, for example, by rotating the gas \cite{Cooper,Sheehy3}. Under rotation, the neutral system behaves like a charged system in a magnetic field. Another way to produce  fictitious magnetic fields is via vortex nucleation \cite{Dalibard}. Also, when a neutral system moves in a suitably designed laser field, it can mimic the dynamics of a charged particle in a magnetic field \cite{Cooper}. We use perturbation theory to solve the Bogoiubov equations and obtain
the wave functions. We consider the various scattering regions of the BCS
regime associated with different energies of the incident particle/hole and
analytically obtain the transmission coefficients and the heat conductivity across the interface. For definite choices of the field, considering various values of the scattering length in the BCS range and taking Andreev reflection into account, we describe how the heat conductivity is affected by the orbital field and the species population imbalance. In particular, we show that the additional heat conductivity due to the orbital field increases with the species imbalance at higher temperatures. Our results indicate how the heat conductivity may be controlled, which is relevant to sensitive magnetic field sensors/regulators at the interface.

\section{Perturbed Bogoliubov equation}
We consider a Fermi gas consisting of two imbalanced species $a$ and $b$, each with mass $m$, effective charge $e$ and chemical potential $\mu_{i}$ ($i=a,b$). (The external trapping potential is assumed to be zero.)  The Hamiltonian operator in the presence of the orbital magnetic field is ($\hbar=1$) \cite{MacDonald}
\begin{equation}
 H_{a,b}=-\frac{1}{2m}(\nabla-ie\bm{A})^{2}-\mu\mp h
\end{equation} 
where $\mu=(\mu_{a}+\mu_{b})/2$ and $h=(\mu_{a}-\mu_{b})/2$ are the average and the species-imbalance chemical potentials, respectively. Treating the vector potential as a perturbation, we may use the first order perturbation theory to solve the Bogoliubov equations \cite{deGennes}. Denoting the particle-like and hole-like solutions by $u_{i}(\bm{r})$ and $v_{i}(\bm{r})$, respectively, we have
\begin{equation}
u_{i}=u^{(0)}_{i}+u^{(1)}_{i}
,\ \ v_{i}=v^{(0)}_{i}+v^{(1)}_{i}, \ \ \Delta=\Delta^{(0)}+\Delta^{(1)}
\end{equation} 
where $u^{(0)}_{i}(\bm{r})$ and $v^{(0)}_{i}(\bm{r})$ are the solutions to the unperturbed ($\mathbf{A}=0$)
Bogoliubov equations and $\Delta(\bm{r})$ is the pair
potential. We use the approximation that $\Delta^{(0)}$ is independent of $\bm{r}$
and by suitable choice of $\bm{A}$ (see later) the contribution of its component perpendicular to the SF boundary can be made negligible. Thus, and using the London gauge ($\nabla \cdot \bm{A}=0$), one finds that
$\Delta^{(1)}\approx 0$ \cite{Ketterson}. The Bogoliubov equations then read:
\begin{eqnarray}
\left(\frac{1}{2m}\nabla^{2}+E+\mu+h\right)u^{(1)}_{a}
-\Delta v^{(1)}_{b}=\frac{ie}{m} \bm{A} \cdot \nabla u^{(0)}_{a}\nonumber \\
\left(-\frac{1}{2m}\nabla^{2}+E-\mu+h\right)v^{(1)}_{b}-\Delta u^{(1)}_{a}=\frac{ie}{m} \bm{A} \cdot \nabla v^{(0)}_{b}.\label{bogo}
\end{eqnarray}
The second set of Bogoliubov equations is obtained by simply interchanging $a$ and $b$. It is clear from these equations that the incoming
$a$-particles are coupled to $b$-holes and vice versa. To proceed, let us take the N-SF interface to be in the $x=0$ plane. Introducing the superscript $s$ ($n$) for the solutions in the SF (N) phase, we expand $u_i^{(n/s)}, v_i^{(n/s)}$ in terms of the unperturbed eigenfunctions. The latter are
$\phi_{\bm{k}(q)}^{\pm}(\bm{r})=\exp[i(\bm{k}_\parallel \cdot \bm{r}\pm k_{(q)}x)]$ for the N phase, and $\phi^{\pm(q)}_{\bm{k}}(\bm{r})=\exp[i(\bm{k}_\parallel \cdot \bm{r}\pm k^{(q)}x)]$ for the SF phase, where $q=p,h$ refers to particle, hole and $\bm{k}_\parallel$ ($k^{(q)},k_{(q)}$) denotes the component of the wave vector $\bm{k}$ parallel (perpendicular) to the interface. Notice that $q$ appears as a subscript (superscript) for the N (SF) phase throughout our notation. From the unperturbed Bogoliubov equations we have the usual relations
\begin{equation}
 k_{(h)}=\sqrt{k^{2}_{(p)}-4m(E+h)}, \ \ \
k^{(p,h)}=\sqrt{k^{2}_{(p)}-2m\xi_{\mp}} \label{k}
 \end{equation}
 where $\xi_{\pm}=E+h\pm\sqrt{(E+h)^{2}-\Delta^{2}}$. Thus, for the N phase, we write 
\begin{eqnarray}
u^{(n)}_{a\bm{k}}(\bm{r})=\sum_{\sigma=\pm} U^{\sigma}_{\bm{k}(p)}\phi^\sigma_{\bm{k}(p)}(\bm{r})+ \text{perturbation terms (PT)}\nonumber \\
v^{(n)}_{b\bm{k}}(\bm{r})=\sum_{\sigma=\pm} V^{\sigma}_{\bm{k}(h)}\phi^\sigma_{\bm{k}(h)}(\bm{r})+\text{perturbation terms (PT)}. \label{N}
\end{eqnarray}
As for the SF phase,
\begin{eqnarray}
u^{(s)^{(0)}}_{a\bm{k}}(\bm{r})=\sum_{q,\sigma}U^{\sigma(q)}_{\bm{k}}\phi^{\sigma (q)}_{\bm{k}}(\bm{r}), \ \ \
v^{(s)^{(0)}}_{b\bm{k}}(\bm{r})=\sum_{q,\sigma}V^{\sigma(q)}_{\bm{k}}\phi^{\sigma(q)}_{\bm{k}}(\bm{r}) \nonumber \\ 
u^{(s)^{(1)}}_{a\bm{k}}(\bm{r})=\sum_{\bm{k}',q,\sigma}a^{\sigma(q)}_{\bm{k}\bm{k}'}\phi^{\sigma(q)}_{\bm{k}'}(\bm{r}) \ \ \
v^{(s)^{(1)}}_{b\bm{k}}(\bm{r})=\sum_{\bm{k}',q,\sigma}b^{\sigma(q)}_{\bm{k}\bm{k}'}\phi^{\sigma(q)}_{\bm{k}'}(\bm{r})\label{expn}
\end{eqnarray}
where $V^{\sigma(p)}_{\bm{k}}=B U^{\sigma(p)}_{\bm{k}}$ and $V^{\sigma(h)}_{\bm{k}}=B^{-1} U^{\sigma(h)}_{\bm{k}}$ with $B=\xi_-/\Delta$.
Substituting (\ref{expn}) into (\ref{bogo}), we can use the orthonormality of $\phi$'s to extract the coefficients $a^{\sigma(q)}_{\bm{k}\bm{k}'}$ and $b^{\sigma(q)}_{\bm{k}\bm{k}'}$. We obtain, after some calculations,
\begin{eqnarray}
a^{\sigma(q)}_{\bm{k}\bm{k}'}= \frac{M^{\sigma(q)}_{\bm{k}\bm{k}'}} {f_+^{(q)} f_-^{(q)}-\Delta^2}(f_-^{(q)} U^{\sigma(q)}_{\bm{k}}+\Delta V^{\sigma(q)}_{\bm{k}})\nonumber \\
b^{\sigma(q)}_{\bm{k}\bm{k}'}= \frac{M^{\sigma(q)}_{\bm{k}\bm{k}'}} {f_+^{(q)} f_-^{(q)}-\Delta^{2}}(\Delta U^{\sigma(q)}_{\bm{k}}+f_+^{(q)} V^{\sigma(q)}_{\bm{k}}) \label{a}  
\end{eqnarray}
where
\begin{eqnarray}
M^{\sigma(q)}_{\bm{k}\bm{k}'}=\frac{ie}{2m} \int d^3x \:[\phi^{\sigma(q)\ast}_{\bm{k}'}\nabla \phi^{\sigma (q)}_{\bm{k}}-\phi^{\sigma(q)}_{\bm{k}}\nabla \phi^{\sigma (q)\star}_{\bm{k}'}]\cdot \bm{A} \nonumber \\
f_\pm^{(q)}=E+h\pm[\mu-\frac{1}{2m}(k_\parallel^2+k^{(q)^2})].\ \ \ \ \ \ \ \ 
\end{eqnarray}
In our calculations we have naturally supposed that for $\sigma \neq \sigma'$,
\begin{equation}
\int d^3x \:[\phi^{\sigma(q)\ast}_{\bm{k}'}\nabla \phi^{\sigma' (q)}_{\bm{k}}-\phi^{\sigma'(q)}_{\bm{k}}\nabla \phi^{\sigma (q)\star}_{\bm{k}'}]\cdot \bm{A} =0.
\end{equation}
(The integrand, then, does not pertain to a single entity (particle/hole) and is thus irrelevant as far as the transmission of particles/holes is concerned.) The amplitudes $U$ and $V$ are to be determined by matching the wave functions and their derivatives at $x=0$, of course. This is deferred to the next section. 

\section{Transmission coefficients}
 
Consider an incoming $a$-particle from the N side with energy $E>\xi_{(p)}-\mu_{a}$, where $\xi_{(p)}\equiv k^{2}_{(p)}/2m$. Because there is no correction to the energy to lowest order in $\bm{A}$ \cite{Ketterson}, in view of (\ref{k}), there are different scattering regimes involved in the $E-\xi_{(p)}$ plane. These regimes are obtained by the crossover of either of the $
k^{(p,h)}_{(h)}$ or $\bm{k}_\parallel$ from real to imaginary values (or vice versa), signifying a change in the scattering mechanism as described below \cite{Lazarides1,Lazarides2}: 

\noindent i)  For $E<\Delta-h$, the incoming particle has insufficient energy to
 excite the SF side. In this case we have total reflection.
 
\noindent ii)  For $E>\Delta-h$ and $\xi_{(p)}< \xi_-$, the incoming particle is reflected.

\noindent iii) For $ E>\Delta-h$ and $\xi_-<\xi_{(p)}<\xi_+$, only quasi-particles are excited in the SF.

\noindent iv) For $\xi_+<\xi_{(p)}<2(E+h)$, particle-like and hole-like excitations both occur in the SF side but
Andreev reflection \cite{Griffin,Andreev1,Andreev2} is forbidden.

\noindent v) For $2(E+h)<\xi_{(p)}<E+\mu_{a}$, we have particle-like and hole-like excitations as well as normal and Andreev reflections.

Because in regions i and ii the particle has insufficient energy to excite the SF side, the
transmission coefficients in these regions vanish. Moreover, our focus is on energies slightly above the transmission threshold ($E\approx \Delta-h$), because we are considering low temperatures (see later). Therefore, we concentrate on regions iv and v. We proceed to obtain $U$ and $V$ by matching the wave functions and their derivatives. For region iv we have
\begin{eqnarray}
\left( \begin{array}{cccc} 
1+\sum_{\bm{k}'}X_{\bm{k}\bm{k}'}& 1+\sum_{\bm{k}'}Y_{\bm{k}\bm{k}'} & 0 & 0 \\
B^{-1}(1+\sum_{\bm{k}'}Y'_{\bm{k}\bm{k}'}) & B(1+\sum_{\bm{k}'}X'_{\bm{k}\bm{k}'}) & 0 & 0\\
 -k^{(h)}-\sum_{\bm{k}'}k'^{(h)}X_{\bm{k}\bm{k}'}& k^{(p)}+\sum_{\bm{k}'}k'^{(p)}Y_{\bm{k}\bm{k}'} & 0 & 0\\
-B^{-1}(1+\sum_{\bm{k}'}k'^{(h)}Y'_{\bm{k}\bm{k}'}) & B(k^{(p)}+\sum_{\bm{k}'}k'^{(p)}X'_{\bm{k}\bm{k}'}) & 0 & 0\\
\end{array} 
\right)
\left( \begin{array}{cccc}
U^{-(h)}_{\bm{k}} \\ U^{+(p)}_{\bm{k}} \\0 \\0 \\
\end{array} 
\right)= \nonumber \\
\left( \begin{array}{cccc}
U^+_{\bm{k}(p)}+U^-_{\bm{k}(p)}+PT\\ V^-_{\bm{k}(h)}\\
k_{(p)}(U^+_{\bm{k}(p)}-U^-_{\bm{k}(p)})+PT\\  -k_{(h)}V^-_{\bm{k}(h)}\\
\end{array} 
\right) \label{matrix}
\end{eqnarray}
 where
 \begin{eqnarray}
X_{\bm{k}\bm{k}'}=M^{-(h)}_{\bm{k}\bm{k}'}\frac{f^{(h)}_-+\Delta B^{-1}}{f^{(h)}_+f^{(h)}_--\Delta^2}, \ \
Y_{\bm{k}\bm{k}'}=M^{+(p)}_{\bm{k}\bm{k}'}\frac{f^{(p)}_-+\Delta B}{f^{(p)}_+f^{(p)}_--\Delta^2} \nonumber\\
X'_{\bm{k}\bm{k}'}=M^{+(p)}_{\bm{k}\bm{k}'}\frac{f^{(p)}_++\Delta B^{-1}}{f^{(p)}_+f^{(p)}_--\Delta^2}, \ \
Y'_{\bm{k}\bm{k}'}=M^{-(h)}_{\bm{k}\bm{k}'}\frac{f^{(h)}_++\Delta B}{f^{(h)}_+f^{(h)}_--\Delta^2}.
\end{eqnarray}
Similar equations are obtained for region v. These equations yield the scattering amplitudes $U^{\sigma (q)}_{\bm{k}}$ for general ${\bm{A}}$. Denoting the $x$-component of the current density by $j_x$, the transmission coefficient is given by $W=j^{\text{T}}_x/j^{\text{I}}_x$, where the superscripts T and I refer to the transmitted
and incident quasi-particle current densities, respectively. The general form of $\bm{j}$ (for channel $\alpha$) is 
\begin{eqnarray}
\bm{j}_\alpha (\bm{r})=-\frac{i}{2m}[u^\star_a(\bm{r})\nabla u_a(\bm{r})-u_a(\bm{r})\nabla u^\star_a(\bm{r})-v^\star_b(\bm{r})\nabla v_b(\bm{r})+v_b(\bm{r})\nabla v^\star_b(\bm{r})] \nonumber \\
-\frac{e}{m}[u^\star_a(\bm{r})u_a(\bm{r})+v^\star_b(\bm{r})v_b(\bm{r})]\bm{A}.
\end{eqnarray}
Using (\ref{N}) and (\ref{expn}) we obtain, to first oder in the perturbation, 
\begin{equation}
W=W^{(0)}+\frac{1}{k_{(p)}} \sum_{\bm{k}'}[\eta^{(1)}_{\bm{k}\bm{k}'} (k'^{(h)}-k^{(p)})+\eta^{(2)}_{\bm{k}\bm{k}'} (k'^{(p)}-k^{(h)})+\eta^{(3)}_{\bm{k}\bm{k}'}(k'^{(h)}+k^{(h)})+\eta^{(4)}_{\bm{k}\bm{k}'} (k'^{(p)}+k^{(p)})] \label{W}
\end{equation}
where $W^{(0)}$ is the transmission coefficient in the absence of the vector potential and
\begin{eqnarray}
\eta^{(1)}_{\bm{k}\bm{k}'}=\frac{U^{+(p)\star}_{\bm{k}} U^{-(h)}_{\bm{k}}}{U^{+^2}_{\bm{k}(p)}}\:(Y'_{\bm{k}\bm{k}'}-X_{\bm{k}\bm{k}'}), \ \ \eta^{(2)}_{\bm{k}\bm{k}'}=\frac{U^{+(p)}_{\bm{k}} U^{-(h)\star}_{\bm{k}}}{U^{+^2}_{\bm{k}(p)}}\:(Y_{\bm{k}\bm{k}'}-X'_{\bm{k}\bm{k}'})\nonumber \\
\eta^{(3)}_{\bm{k}\bm{k}'}=\frac{ U^{-(h)^2}_{\bm{k}}}{U^{+^2}_{\bm{k}(p)}}\:(B^{-2}Y'_{\bm{k}\bm{k}'}-X_{\bm{k}\bm{k}'}), \ \ \eta^{(4)}_{\bm{k}\bm{k}'}=\frac{U^{+(p)^2}_{\bm{k}}}{U^{+^2}_{\bm{k}(p)}}\:(Y_{\bm{k}\bm{k}'}-B^2 X'_{\bm{k}\bm{k}'}). 
\end{eqnarray}
In the computations (which are performed analytically throughout) we use
\begin{equation}
\sum_{\bm{k}'} \rightarrow \int \frac{d^3\bm{k}'}{(2\pi)^3}=\frac{m}{4\pi^2} \int dE'_\alpha \int dk'_{(p)}
\end{equation}
as we are considering the $\alpha$ channel. Note that only energies slightly above the transmission threshold contribute to the first integral, which affects the region of integration. We also need to make a definite choice for $\bm{A}$. Equation (\ref{matrix}) then yields very complicated analytical expressions for the scattering amplitudes that reduce to the results of \cite{Lazarides2} in the absence of the vector potential. Substituting these in (\ref{a}) gives the coefficients $a^{\sigma(q)}_{\bm{k}\bm{k}'}$ and $b^{\sigma(q)}_{\bm{k}\bm{k}'}$.

We take $\bm{A}=\frac{c}{(x-x_{0})^{2}}\hat{\bm{i}}+\frac{cy}{(x-x_{0})^{3}}\hat{\bm{j}}+\frac{cz}{(x-x_{0})^{3}}\hat{\bm{k}}$, where $c$ and $x_0\neq 0$ are constants. (This particular choice renders analytical computation possible.) Equation (\ref{W}) leads to 
 \begin{equation}
W_\pm=W^{(0)}_\pm+2ec\sqrt{\frac{m\Delta}{\chi}}\: \frac{\chi-\Delta\pm\sqrt{\chi(\chi-\Delta)}}{\chi\mp\sqrt{\chi^2-\Delta^2}}\:(\Delta-h)\sqrt{E-\Delta+h}
 \end{equation}
 where $\chi=\xi_{(p)}-\Delta$ and $W_+$ ($W_-$) refers to region iv (v). $W^{(0)}_\pm$ coincide with the results of \cite{Lazarides2} in the absence of the vector potential ($c=0$), which are also proportional to $\sqrt{E-\Delta+h}$. We remark that only a tiny portion of region iv (corresponding to $\xi_{(p)}$ slightly above $2\Delta$ and, of course, $E$ slightly above $\Delta-h$) contributes to $W_+$. Similarly for holes, we obtain results for the region v with $\xi_{(h)}=k^{2}_{(h)}/2m$. This is because from our calculations, holes incident on the interface can only transmit in region v. 

Analogous procedure applied to incoming $b$-particles and $a$-holes (the $\beta$ channel) yields identical results to the above but with $h\rightarrow -h$.

\section{Heat conductivity}
For each channel, the heat conductivity is given by
\begin{equation} \kappa=\frac{m}{4\pi^{2}}\frac{\partial}{\partial
T}\sum_{q=p,h} \int d\xi_{(q)}\int
dE E f(E)W(E,\xi_{(q)}) \label{cond}
 \end{equation}
where $f(E)$ is Fermi-Dirac distribution, which, for sufficiently low temperatures $T<<\Delta$ ($k_{\text{B}}=1$),  reduces to $e^{-(\Delta-h)/T}$ (up to a proportionality constant). This implies that the relevant range of energies is only slightly above the threshold $E=\Delta-h$, as mentioned before. 
 
Using (\ref{cond}) we find $\kappa=\kappa^{(0)}+\kappa^{(1)}=(\kappa^{(0)}_{\alpha}+\kappa^{(0)}_{\beta})+(\kappa^{(1)}_{\alpha}+\kappa^{(1)}_{\beta})$,
where $\kappa^{(0)}$ is the heat conductivity in the absence of the external field and $\kappa^{(1)}$ is the additional term  due to the vector potential in the SF side. The effect of the vector potential in the N side on $\kappa^{(1)}$ is of second order and, therefore, negligible.  Its effect on $\kappa^{(0)}$ is to add a small term proportional to $\kappa^{(0)}$ itself. This does not change the temperature dependence of $\kappa^{(0)}$ and only causes a small shift in the $\kappa^{(0)}-T$ curve (figure 3 of reference \cite{Lazarides1}), which we neglect. Also, in our calculations we have neglected the energy carried by the $\beta$-channel because it is quite lower than that carried by the $\alpha$-channel. Our final expression for $\kappa^{(1)}$ is
\begin{equation}
\kappa^{(1)}=12\surd 2 ec\Delta (\Delta-h)^{9/2}m^{3/2}(\sqrt{2\Delta}-\sqrt{\mu+\Delta}+\frac{1}{\sqrt{\mu-\Delta}})\frac{e^{-(\Delta-h)/T} }{{\sqrt{\Delta T}}}.
\end{equation}

\begin{figure}
\includegraphics{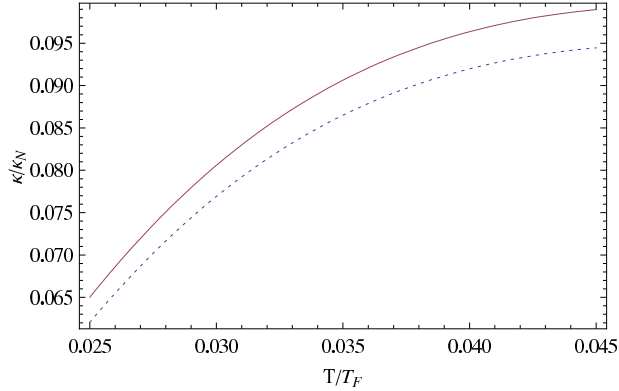}
\caption{(Color online) Heat conductivity versus temperature across the interface in the BCS
regime for $c=0.45$. The dotted line represents the heat conductivity in the absence of the
vector potential.} \label{fig1}
\end{figure}

\begin{figure}
\includegraphics{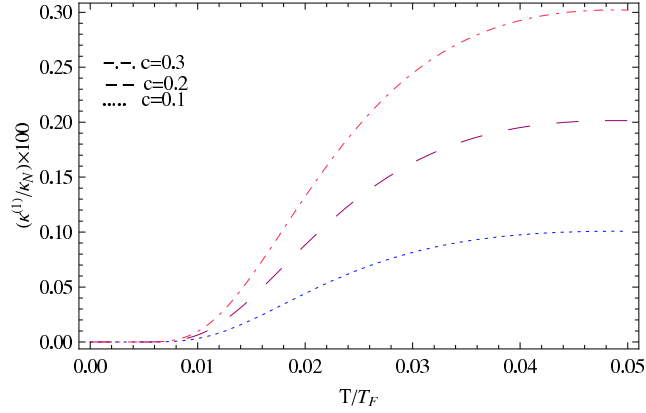}
\caption{(Color online) Additional heat conductivity versus temperature across the interface in the BCS regime for various values of $c$.} \label{fig2}
\end{figure}

\begin{figure}
\includegraphics{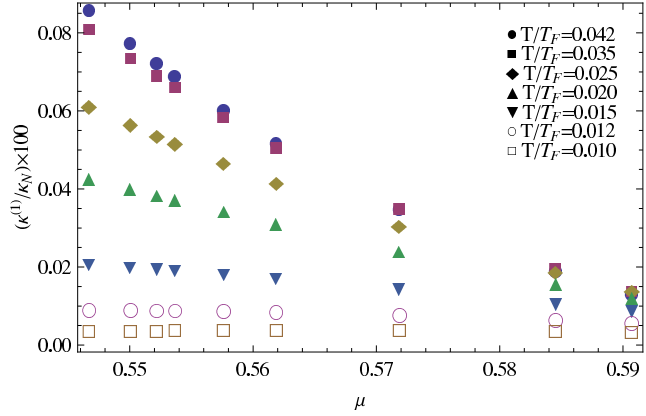}
\caption{(Color online) Additional heat conductivity versus average chemical potential across the
interface in the BCS regime for $c=0.1$.} \label{fig3}
\end{figure}

\begin{figure}
\includegraphics{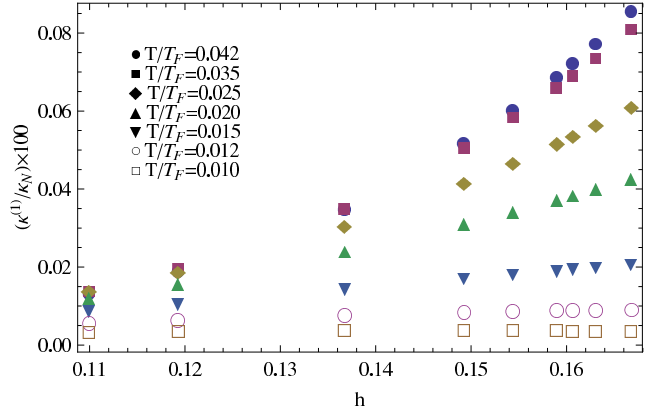}
\caption{(Color online) Additional heat conductivity versus imbalance chemical potential across the interface in BCS
regime for $c=0.01$.}
\label{fig4}
\end{figure}

By fixing the values of $h$ and $\mu$ using the gap and number equations \cite{Lazarides2,Shin1}, we plot in FIG. \ref{fig1}, $\kappa/\kappa_{\text{N}}$ ($\kappa_{\text{N}}=2m\mu T/\pi^2$, heat conductivity in the
normal phase) as a function of $T/T_{\text{F}}$ in the BCS regime ($\Delta=0.25T_{F}$ and $(k_{\text{F}}a)^{-1}=-0.44$). 
The effect of the vector potential is simply a shift of the heat conductivity curve. We have plotted the figure in the range $0.030T_{\text{F}}<T<0.045T_{\text{F}}$, where the shift is more pronounced. Also in FIG. \ref{fig2} we plot $\kappa^{(1)}/\kappa_{\text{N}}$ versus $T/T_{\text{F}}$, for various values of $c$.
As seen, at fixed $T$, the larger the value of $c$, the larger is the heat conductivity. However, at temperatures roughly below $0.01 T_{\text{F}}$ the effect of the vector potential is negligible. Moreover, the curves show an increasing heat conductivity, with a rate that decreases substantially after around $T=0.03 T_{\text{F}}$. FIG. \ref{fig3} shows the dependence of $\kappa^{(1)}/\kappa_{\text{N}}$ on $\mu$ at fixed temperature for various values of $T/T_{\text{F}}$.  
To obtain this plot, we have first fixed $\Delta a^{2}$ ($a$, the scattering length). Using the gap and number equations, we then fix $k_{\text{F}}a$ (such that it lies in the BCS range) which fixes the values of $\mu$, $h$ and $\Delta$. As seen, in the specified temperature range, $\kappa^{(1)}$ does not change significantly with $\mu$ at lower temperatures but decreases as the temperature is raised. Notice also the (almost) collapse of the data points, showing that the graph is roughly independent of the temperature as long it is high enough. We also show, in FIG. \ref{fig4}, the dependence of $\kappa^{(1)}/\kappa_{\text{N}}$ on $h$ at fixed temperature for various values of $T/T_{\text{F}}$.
Again, $\kappa^{(1)}$ does not change significantly at lower temperatures but now increases (in contrast to the previous figure) as the temperature is raised. Similarly, the graph is roughly independent of the temperature for high enough values. Since $h$ can be controlled by the species population imbalance, the graph shows how the effect of the weak magnetic field on the heat conductivity may be controlled.

\begin{acknowledgments}
One of the authors (R.A.) is grateful to A.J. Leggett, C.J. Pethick, H. Shimahara and D.E. Sheehy for valuable guide.
\end{acknowledgments}

\end{document}